\documentclass{article}
\usepackage[utf8]{inputenc}
\usepackage{amsfonts}
\usepackage{amsmath}
\usepackage{MnSymbol}
\usepackage{wasysym}
\usepackage{tikz}
\usepackage{tcolorbox}
\usepackage{multirow}
\usepackage{hhline}
\usepackage{xcolor}
\usepackage{colortbl}
\usepackage{graphicx}
\usepackage{booktabs}
\usepackage{epstopdf}
\usepackage{color}
\usepackage{float}
\usepackage[style=authoryear,sorting=nyt,natbib=true,backend=bibtex]{biblatex}
\usepackage{pdfpages}
\usepackage{pgf,tikz,pgfplots}
\pgfplotsset{compat=1.15}
\usepackage{mathrsfs}
\usetikzlibrary{arrows}
\usepackage{tikz-3dplot}
\usepackage[utf8]{inputenc}
\usepackage{graphicx}
\newtheorem{theorem}{Theorem}

\newtheorem{axiom}{Axiom}

\newtheorem{example}[theorem]{Example}

\newtheorem{proposition}{Proposition}

\usepackage{tikz}
\usepackage{verbatim}
\usetikzlibrary{arrows,positioning} 
\tikzset{
	>=stealth',
	punkt/.style={
		text width=7.5em,
		minimum height=1em,
		text centered},
	pil/.style={
		->,
		thick,
		shorten <=2pt,
		shorten >=2pt,}
}

\begin{document}

	\title{A Graph-based Similarity Function for CBDT: Acquiring and Using New Information}
	\author{Federico Contiggiani \\
		Universidad Nacional de Río Negro\\ 
		Instituto de Investigación en Políticas Públicas y Gobierno \\
		fcontiggiani@unrn.edu.ar (corresponding author) \and Fernando Delbianco \\
		Departamento de Economía, Universidad Nacional del Sur\\Instituto de Matemática de Bahía Blanca\\
		CONICET - Universidad Nacional del Sur \\
		fernando.delbianco@uns.edu.ar \and Fernando Tohm\'e\\
		Departamento de Economía, Universidad Nacional del Sur\\Instituto de Matemática de Bahía Blanca\\
		CONICET - Universidad Nacional del Sur \\
		ftohme@criba.edu.ar }
	\maketitle

\begin{abstract}
One of the consequences of persistent technological change is that it force individuals to make decisions under extreme uncertainty. This means that traditional decision-making frameworks cannot be applied. To address this issue we introduce a variant of Case-Based Decision Theory, in which the solution to a problem obtains in terms of the distance to previous problems. We formalize this by defining a space based on an orthogonal basis of {\em features} of problems. We show how this framework evolves upon the acquisition of new information, namely features or values of them arising in new problems. We discuss how this can be useful to evaluate decisions based on not yet existing data.
\end{abstract}



\bigskip $^{{}}$

\section{Introduction}\label{sec1}

One of the main characteristics of accelerated technological change is the continuous appearance of new products on the market. But other than in a very few cases, these products are variants of already existing ones to which new features have been added. For consumers, this raises a high degree of uncertainty about those additions. On one hand, there is the problem of how to decide on buying a brand or another of a recently launched product. The lack of familiarity with the new features hampers the ability of making fully informed decisions. On the other hand, this uncertainty is relevant for the decision on when to buy technological gadgets, either now or in the near future. 

Such problems can be addressed with cognitive models of human decision making conveying relevant insights on the psychological and computational process that operate when the individual evaluates alternatives and makes a decision \parencite{Lipman1995}. Behavioral approaches to individual decision making lend support to the metaphor that human choices might be well represented as an information processor that is influenced by elements of the decision context  \parencite{Rubinstein1998}. 

Information processing is critically dependent on the accessibility to the right sources and the accuracy of data. This is even more relevant when this information is required for decision-making under uncertainty. Some important questions in this respect are, {\it What information is needed to solve a problem?}, {\it How can the relevant factors be weighted up in the search of solutions?} and {\it How can new aspects of the problem be learned or inferred?} Recent contributions show that these questions are still open and call for new concepts and methodologies \parencite{newell2008cognitive, lipshitz2001taking}.

Our take on this subject is based on a particular model of Case-Based Decision Theory (CBDT). This approach assumes that the similarity with previous problems is critical to the way a new problem is addressed. Similar approaches in the judgment and decision-making (JDM) literature refer to \textit{exemplar} models for the study of categorization strategies \parencite{karlsson2008exemplar}, cue-based inferences elaboration and example-based reasoning \parencite{platzer2013rule}. In addition  there are methodological questions common to both CBDT and Naturalistic Decision Making \parencite{lipshitz2001taking}.

The canonical approach to individual decision making under uncertainty is Expected Utility Theory (EUT), which assumes that rational agents enumerate all the possible states of the world and the corresponding consequences associated with them. Furthermore, they are able to assess the probabilities of all possible relevant states of the world. In the real world, agents are usually unable to fully describe the class of states of the world, mainly because of the complexity involved in this task.

Instead, individuals facing decision problems make their decisions based on their previous experience, searching in their memories to recall what they did in similar situations, in order to assess the convenience of choosing the same actions as in the past. This intuition is captured by CBDT, as presented by \cite{Gilboa1995,Gilboa2001,Gilboa2003} and \cite{Matsui2000}. In CBDT the preferences of a decision-maker over actions to be exerted to solve a new problem result from the history of previous problems faced by her, which are stored in her memory. 

A few assumptions ensure that those preferences can be captured in an utility function over actions, defined as the sum of values of adopting them on previous problems. Each of these values is weighted up by the similarity between the previous and current problem. Similarities are so crucial in this framework that we can identify an agent at a new problem with the ensuing similarity function. The original formulation of CBDT does not assume any particular shape of this function. But in other areas, analogous relations have been characterized \parencite{Tversky1977,Lipman1995,Rubinstein1998,Mullainathan2002,Kahneman2003,Johnson2014}.

According to this, we can define similarity in terms of a ``distance'' between problems, which can obtained, in turn, comparing the features characterizing them. The closer two problems are in terms of such distance, the higher is their similarity. In our framework we construct a graph embedded in a real metric space. Each node is identified with a problem and the edges connecting pairs of problems have an attached weight, namely the distance between them. When a new problem arises, it becomes a point in the metric space. The problems in the graph that are \emph{closer} to it become relevant to determine the actions to be carried out. Learning amounts to reconfigure the metric space and the embedded graph upon the acquisition of new pieces of information.

The metric space in which the graph of problem is embedded is defined by the \emph{features} that constitute the dimensions on which the similarities of problems are evaluated. This is close to the idea that decisions are made in terms of those made previously by peers \parencite{Li2017}. Furthermore, we evaluate how decisions are made in the presence of new values of the features or in the face of the realization of the relevance of new features. Furthermore, we consider how the anticipation of future new values or features affect the decisions.

In this paper we present the formal aspects of this decision-making framework, starting with a thorough description of CBDT. We add an extra axiom that ensures the characterization of similarity in terms of the distance between problems. Then we introduce the metric space capturing this concept of distance and analyze how it evolves in response to new information.

Our main concern is with the application of these insights to decision-making involving entities with previously unknown features or, worst yet, with future features with unforeseen aspects. The literature on technological adoption has discussed similar problems, emphasizing on how individuals analyze the different factors involved, and using previous experiences (not necessarily their own one) to make their decisions. There is a large literature that looks for the factors and dimensions by which a gadget or technological device is selected by a consumer. CBDT is especially useful in contextualizing this decision process, due to the intrinsic characteristics of the technology, which make it a continuous learning-and-choosing process \cite{bhui2018case}. \cite{davis1989perceived} elaborates a scheme illustratating how external variables such as the perception of ease of use and usefulness, may or may not lead to the use or adoption of new technologies. \cite{basoglu2017will} presents a non-exhaustive list of additional factors listed by various authors, such as enjoyment (\cite{ingham2015shopping}, \cite{li2017factors}), self-efficacy (\cite{ozturk2017understanding}, \cite{chen2014predictors}), peer influence (\cite{dutot2015factors}), external influence (\cite{chung2014job}), risk (\cite{natarajan2017understanding}, \cite{yoo2015knowing}), attitude (\cite{basoglu2012determining}, \cite{daim2014service}), usefulness (\cite{topacan2010exploring}, \cite{basoglu2007organizational}; \cite{basoglu2009decision}, \cite{daim2010critical}, and \cite{tanoglu2010exploring}), ease of use (\cite{seneler2009interface} and \cite{seneler2010empirical}), anxiety (\cite{kummer2017technology}, \cite{lazuras2016mental}), health concern (\cite{ducey2016predicting}), intention (\cite{daim2014service}, \cite{kargin2009factors}), and complexity (\cite{lazard2014user}; \cite{chin2015investigating}). This brief list helps to illustrate the multi-dimensional nature of the choice process. \cite{dehghani2018will} shows how the definition of a technological gadget varies, indicating how the characteristics defining a watch have evolved in time to end up yielding the concept of smartwatch. This implies that these evolving characteristics of a gadget are subject to a process of active learning, which in turn leads to an increasing probability of new additions. \cite{asadi2019integrated} builds a structural model with a neural network approach, which provides an empirical assessment of this learning-and-choosing process. Finally, another factor making this continuous process of technological adoption so special is the effect of mouth-to-mouth references and on-line reviews, which is particularly fast for technological gadgets. \cite{huang2020impact} assess the empirical validity of this perception in the purchase of household appliances. 

The work is structured as follows. Section 2 presents the formalism of CBDT. Section 3 discusses how to represent CBDT's similarity function in terms of distances on graphs. Section 4 specifies those graphs in spaces determined by \emph{features} of problems. Section 5 studies how the previously presented formalism can be adapted to new information. In turn, Section 6 discusses how agents can learn the rate at which new data arises and use this to make decisions about contexts that are not yet existing. Section 7 concludes.

\section{Definitions and Axioms}\label{sec2}

Gilboa and Schmeidler (1995) assume a finite and nonempty set $P$ which is given as a primitive and contains all the possible \emph{problems} that an agent may face as well as a finite and nonempty set of \emph{actions} $A$. To simplify they assume that all the actions in $A$ are available for any problem $p\in P$. In addition, there is a set $R = \mathbb{R}$ of results. The result of not choosing an action is denoted $r_{0}\in R$ (for simplicity, we assume that $r_0 = 0$). Then, the set of \emph{cases} is $C\equiv P\times A\times R$.\\
The agent is endowed with a \emph{memory} set $M\subseteq C$. Its projection over $%
P $ is called the \textit{history} and is defined by $H\left( M\right)
=\left\{ q\in P|\exists a\in A,r\in R:\left( q,a,r\right) \in M\right\} $.
The set $M$ has the following two properties:

\begin{itemize}
	
	\item for all $q\in H\left( M\right) \wedge a\in A\Longrightarrow
	\exists !r=r_{M}\left( q,a\right) :\left( q,a,r\right) \in M$;
	
	\item for all $q\in H\left( M\right) \Longrightarrow \exists !a\in
	A:r_{M}\left( q,a\right) \neq r_{0}$.
\end{itemize}

The first condition indicates that for every problem $q$ in the history and an action $a$ there exists a unique result of applying this action to solve $q$. The second condition states that for each problem $q$ in the history, there exists a unique action $a$ that yields a non-void result. These two conditions together ensure that for each pair of cases in $H$, $\left(q,a,r\right)$ and $\left(q^{\prime}, a^{\prime}, r^{\prime}\right)$, $q \neq q^{\prime}$ and $a \neq a^{\prime}$.

The agent makes a decision based both on the utility of the results of a
given action, and a \emph{similarity} function $s$ which assigns nonnegative values
to pairs of problems. In this way, when the agent faces a new problem $p$,
he selects an act $a$ that maximizes the following expression:

$$U\left( a\right) =U_{p,M}\left( a\right) =\underset{\left( q,a,r\right) \in
	M}{\dsum }s\left( p,q\right) u\left( r\right)$$

\noindent where $u: R \rightarrow \mathbb{R}^{+}$ is the \emph{instantaneous utility} of results.

Gilboa and Schmeidler present an axiomatic system, representing some desirable properties of the similarity function. Furthemore, they show that there exists a unique $s:P^{2}\rightarrow \left[ 0,1\right] $ such that the function $U\left( \cdot \right) $ is the representation of a preference relation $%
\succeq _{p,H}$, on the class of actions, where $p$ is a new problem not corresponding to any case in the history. In the statement of the axioms each action $x \in A$ is identified with a vector in $\mathbb{R}^{H}$, where each component of the vector is the result of applying action $x$ on a problem  $q$ for which a case $\left(q,a,r\right)$ exists in the history. We denote with $x(q)$ the result of applying action $x$ to problem $q$. If $x = a$, then $x(q) = r$. Otherwise, $x(q) = r_0$.
\begin{axiom}
	Comparability of Compatible Profiles. For every $p\in P$ and every history $H=H\left( M\right) $, for every $x,y\in \mathbb{R}^{H}$, $x$ and $y$ are compatible if and only if $x\succeq _{p,H}y$\ or $y\succeq _{p,H}x$.
\end{axiom}

\begin{axiom}
	Monotonicity. For every $p$, $H$, $x\geq y$ and $x\ast y=0$ implies that $x\succeq _{p,H}y$.
\end{axiom}

\begin{axiom}
	Continuity. For every $p$, $H$, and $x\in\mathbb{R}^{H}$, the sets $\left\{ y\in\mathbb{R}^{H}|y\succeq _{p,H}x\right\}$ and \newline $\left\{ y\in \mathbb{R}^{H}|x\succeq _{p,H}y\right\}$ are closed (in the standard topology on $\mathbb{R}^{H}$).
\end{axiom}

\begin{axiom}
	Separability. For every $p$, $H$, and $x,y,z,w\in\mathbb{R}^{H}$, if $\left( x+z\right) \ast \left( y+w\right) =0$, $x\succeq _{p,H}y$,
	and $z\succeq _{p,H}w$, then $\left( x+z\right) \succeq _{p,H}\left(
	y+w\right) $.
\end{axiom}

\begin{axiom}
	Similarity Invariance. For every $p,q_{1},q_{2}\in P$ and every two memories 
	$M^{1},M^{2}$\ with $q_{1},q_{2}\in H^{i}\equiv H\left( M^{i}\right) $
	(i=1,2) and $p\notin H^{i}$\ (i=1,2), let $v_{j}^{i}$ stand for the unit
	vector in $\mathbb{R}^{H^{i}}$\ (i=1,2) corresponding to $q_{j}$ (j=1,2) (That is, $v_{j}^{i}$\
	is a vector whose $q_{j}$th component is 1 and its other components are 0).
	Then denoting the symmetric part of $\succeq _{p,H}$ by $\approx _{p,H}$,
	
	$x,y\in \mathbb{R}^{H^{1}},$ $z,w\in\mathbb{R}^{H^{2}},$ $x\approx _{p,H^{1}}y,$ $z\approx _{p,H^{2}}w$\qquad
	
	and
	
	$x+\alpha v_{1}^{1}\approx _{p,H^{1}}y+\beta v_{2}^{1}$
	
	imply that
	
	$z+\alpha v_{1}^{2}\approx _{p,H^{2}}w+\beta v_{2}^{2}$\qquad
	
	whenever the compared profiles are compatible.
\end{axiom}

Given these axioms and using claims proven by Gilboa and Schmeidler, we can show
that the similarity between pairs of problems is representable in terms of a
connected graph $G$ in which each problem in $P$ is assigned a node. The class of nodes of $G$ is $V$ and thus $V\equiv P$. To get there, we need first to consider two extra axioms.
\begin{axiom}
	Symmetric Similarity. For every $r,p,m\in P$ and every three memories $%
	M^{1},M^{2},M^{3}$ with $m,r\in H^{1}\equiv H\left( M^{1}\right) $, $p\notin
	H^{1}$, $m,p\in H^{2}\equiv H\left( M^{2}\right) $, $r\notin H^{2}$, $p,r\in
	H^{3}\equiv H\left( M^{3}\right) $, and $m\notin H^{3}$; let $v_{j}^{i}$\
	stand for the unit vector in $\mathbb{R}^{H^{i}}$\ (i=1,2,3) corresponding to $j$ (j=r,p,m) (That is, is a vector
	whose $j$th component, which is the component associated with the case $j$,
	is 1 and its other components are 0). Then denoting the symmetric part of $\succeq _{p,H}$ by $\approx _{p,H}$,
	
	$x,y\in\mathbb{R}^{H^{1}},$ $z,w\in \mathbb{R}^{H^{2}},$ $l,h\in\mathbb{R}^{H^{3}},$ $x\approx _{p,H^{1}}y,$ $z\approx _{r,H^{2}}w,$ $l\approx
	_{m,H^{3}}h$\qquad \qquad
	
	and
	
	$x+\alpha v_{m}^{1}\approx _{p,H^{1}}y+v_{r}^{1}$, $z+\beta v_{m}^{2}\approx
	_{r,H^{2}}w+v_{p}^{2}$\qquad
	
	imply that
	
	$l+\alpha v_{p}^{3}\approx _{m,H^{3}}h+\beta v_{r}^{3}$\qquad
	
	whenever the compared profiles are compatible.
\end{axiom}

Axiom 6 guarantees that the similarity function is symmetric since it leads us to Gilboa and Schmeidler's necessary and sufficient condition for
symmetric similarity, namely that for all $r,p,m\in P$, $s\left( p,m\right)
s\left( m,r\right) s\left( r,p\right) =s\left( p,r\right) s\left( r,m\right)
s\left( m,p\right) $. More precisely,
\begin{theorem}
	The following two statements are equivalent:
	
	a) Axiom 1 to Axiom 6 hold.
	
	b) There exist a unique and symmetric function $s^{\prime }:P^{2}\rightarrow \mathbb{R}_{+}$\ such that for all $p\in P$, every memory $M$ with $p\notin H\equiv
	H\left( M\right) $ and every compatible $x,y\in	\mathbb{R}^{H}$,\qquad
	
	$x\succeq _{p,H}y\Longleftrightarrow \underset{q\in H}{\dsum }s^{\prime }\left( p,q\right) x\left( q\right) \geq \underset{q\in H}{\dsum }s^{\prime}\left( p,q\right) y\left( q\right) $.
\end{theorem}

Now we introduce another axiom that requires that the
similarity function verifies the following inequality, i.e. $s^{\prime
}\left( p,q\right) \geq s^{\prime }\left( p,m\right) +s^{\prime }\left(
m,q\right) $. Intuitively, this amounts to ask that the ``direct'' similarity between
two problems cannot be lower than the sum of the similarities of these problems
mediated by another one. More precisely:\footnote{Notice that this axiom, as shown through its consequences in Theorem 2, precludes the possibility of, given three problems $p, m$ and $q$, to define a similarity function such that $s(p,q) = s(p,m)= s(m,q)$.}
\begin{axiom}
	Triangular inequality. For every $r,p,m\in P$ and every two memories $M^{1},M^{2}$ with $m,r\in H^{1}\equiv H\left( M^{1}\right) $, $p\notin H^{1}$, $m,p\in H^{2}\equiv H\left( M^{2}\right) $, and $r\notin H^{2}$; let $v_{j}^{i}$ stand for the unit vector in $\mathbb{R}^{H^{i}}$ (i=1,2) corresponding to $j$ (j=r,p,m) (That is, $v_{j}^{i}$ is a vector whose $j$th component, which is the component associated with the case $j$, is 1 and its other components are 0). Then denoting the symmetric part of $\succeq _{p,H}$ by $\approx _{p,H}$,
	
	$x,y\in \mathbb{R}^{H^{1}},$ $z,w\in\mathbb{R}^{H^{2}},$ $x\approx _{p,H^{1}}y,$ $z\approx _{r,H^{2}}w$\qquad
	
	and
	
	$x+\alpha v_{m}^{1}\approx _{p,H^{1}}y+v_{r}^{1}$, $z+\beta v_{p}^{2}\approx
	_{r,H^{2}}w+v_{m}^{2}$\qquad
	
	imply that
	
	$y+\beta v_{r}^{1}+\alpha v_{m}^{1}$ $\preceq _{p,H^{1}}x+v_{m\text{ }}^{1}$\qquad
	
	whenever the compared profiles are compatible.
\end{axiom}

The addition of this axiom leads to:

\begin{theorem}~\label{metric}
	The following two statements are equivalent:
	
	a) Axiom 1 to Axiom 7 hold.
	
	b) There exist a unique and symmetric function $s^{\prime }:P^{2}\rightarrow \mathbb{R}_{+}$\ that verifies that $s^{\prime }\left(
	p,q\right) \geq s^{\prime }\left( p,m\right) +s^{\prime }\left( m,q\right) $, such that for all $p\in P$, every memory $M$ with $p\notin H\equiv H\left(
	M\right) $ and every compatible $x,y\in \mathbb{R}^{H}$,\qquad
	
	$x\succeq _{p,H}y\Longleftrightarrow \underset{q\in H}{\dsum }s^{\prime
	}\left( p,q\right) x\left( q\right) \geq \underset{q\in H}{\dsum }s^{\prime
	}\left( p,q\right) y\left( q\right) $.
\end{theorem}

The proof of both theorems is given in the Appendix.

\section{The Similarity Function Represented by a Graph}\label{sec3}

Given Axioms 1 to 7, and the results obtained in the previous section, we
are now in position to introduce a graph-theoretic version of the similarity
function. For this, consider an agent endowed with a memory set $M$ and a connected graph $G$ with nodes $V=P$. 

We assume that the agent is able to compute a distance $d(p,q)$
between $p,q\in V$, which is defined as the length of the shortest path
joining these two nodes, and since $G$ is connected, this distance is a
metric \parencite{Harary1969}.

In this context, {\it length} is defined as the number of occurrences of edges in
an alternating sequence of nodes and edges (\emph{walk}) between $p$ and $q$
\parencite{Harary1969}. In this way, the agent is able to compare any pair of
problems in the graph. Consider the adjacency matrix $B=\left[b_{pq}\right]$ of $G$ that is the $n\times n$ matrix ($|P| = n$) in which  $b_{pq}=1$ if $%
p$ is adjacent with $q$ in $G$ and $b_{pq}=0$ otherwise. Then, the distance
between $p$ and $q$ for $p\neq q$ is the least integer $l$ for which
the $(p,q)$, entry of $B^{l}$\ is nonzero.

\qquad Now, we are going to consider the following results.

\begin{proposition}~\label{model}
	Consider $S_{G}=\left\langle V,d\right\rangle $, where $V$ is the set of nodes of graph $G$ and $d$ is Harary's minimal
	distance defined on $G$. Then, there exists $\rho :\ell \rightarrow
	S_{G}\cup \mathbb{R}^{+}$, where $\ell $ are all the expressions in the formal language
	in which the CBDT axioms are formulated, such that $M_{G}=\left\langle S_{G} \cup \mathbb{R}^{+},\rho \right\rangle $
	is a model of CBDT+Axiom 6+Axiom 7.
\end{proposition}

\begin{proposition}
	CBDT+Axiom 6+Axiom 7 has only one model up to isomorphism.
\end{proposition}

Given these propositions we show that $S_{G}$ provides a valid
representation of the memory and the similarity function because expressions
that are true of $s$ are true in $G$ with distance $d$; and in addition, $G$ is
compatible with axioms 1 to 7. Furthermore, the representation of $s$
using a graph $G$ is unique up to isomorphism. Finally, all the expressions
on utilities are interpreted as arithmetical claims in $\mathbb{R}^{+}$. The proofs of these propositions can be found in the Appendix.

In other words, these propositions show that there exists a unique graph that represents the similarity function. This is an important addition to the usual presentation of CBDT, since the similarity function is derived there from the preferences over outcomes, while here it is furthermore associated to the shortest path distance on a graph. 

We will now consider a concrete definition of graph $G$ and derive an explicit specification of $s$.

\section{A Specification of the Similarity Function}\label{sec4}

In order to present a concrete specification of the results of the previous section we need to define a graph $G$ in which the nodes correspond to problems. One way to do that is by assuming that each problem can be identified to a point in a \emph{space of features} $\mathcal{F}$. More precisely, since we intend to find a way of maximizing a preferential order $\succeq_{p,H}$, we are only interested in a space that represents the features of the problems corresponding to the finite number of cases in $H$. By the conditions on $H$, all these problems are different, and thus each will yield a different point in the space of features.

Suppose that a problem $q \in H(M)$ is defined by $j_q$ features.\footnote{If the problem amounts to, say, choosing a smartphone, we can consider that the relevant features are the price, the brand, the size of its memory, the quality of the camera, etc. While large, the number of features at the moment of making a decision, is finite.}. Thus, the total number of features to consider is $J=\sum_{q \in H(M)} j_q$. While it is clear that some features are correlated to others, we will assume without much loss of generality that the features are independent. This is the case when $J$ is the minimal number of features needed to describe any problem $q \in H(M)$. This simplification allows us to consider that the $J$ dimensions of the space of features are orthogonal.

The remaining question is what spaces correspond to the different features. While some features admit discrete values others require continuous ones. Since we intend to use Harary's distance, we need to be able to define the \emph{adjacency} between problems, and thus the range of values of each feature has to be discrete. We assume thus that we identify each feature $f_j$ with its discrete range.\footnote{In the case of a continuous-valued feature this means that we determine a finite partition of its range. In practical applications the partition would arise from the application of methods like, for instance, CART.} That is, $f_j = \{f_1^j, \ldots, f_{\kappa_{f_j}}^j\}$, where $\kappa_{f_j}$ is the (finite) number of possible values of $f_j$. Notice that this means that $f_j$ is a linearly ordered set. Then, the entire space of features is $\mathcal{F}= \prod_{j=1}^{J} f_j$.

$G$ will consist of the points $H(M) \subseteq \mathcal{F}$. We take then as edges all the linear $L^1$ (``taxicab'' or ``Manhattan'') segments between the points in $\mathcal{F}$, with the proviso that the distance between two consecutive values in a feature, say $f^j_k$ and $f^j_{k+1}$ is $1$. Harary's procedure gives us the \emph{minimal distance} between points on $\mathcal{F}$. Given two problems $q, r \in \mathcal{F}$, the ensuing distance according to this procedure is $d(q,r)$. In turn, the maximal distance between any pair of problems in $H(M)$ (the so called ``diameter'' of graph $G$) is denoted $D_M$.

Then, given two problems $p \notin H(M)$ and $q \in H(M)$, we take $\bar{d}(p,q) = \frac{d(p,q)}{D_M}$
and define 

$$s(p,q) \ = \ 1 - \bar{d}(p,q)$$

It is easy to see that:

\begin{proposition}
	$s(p,q)$ satisfies the conditions of the similarity function of Theorem~\ref{metric}. Furthermore, for any statement about the similarity between any pair of problems $p$ and
	$q$, $\Lambda(p,q)$, $\rho(\Lambda(p,q)) = s(p,q)$, where $\rho$ is as defined in Proposition~\ref{model}.
\end{proposition}

\begin{figure}[h!]\centering
	\includegraphics[scale=1.05]{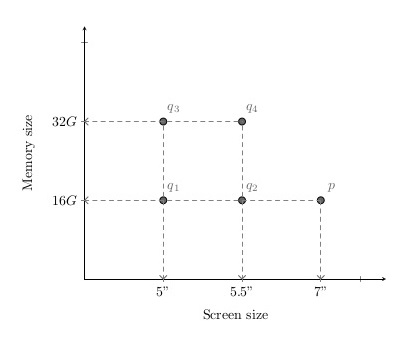}
	\caption{Graph representation of CBDT.} \label{fig:F1}
\end{figure}

\begin{example}~\label{ex1}
	Consider the following context:
	
	$$H(M_{\mathcal{F}_0}) =\left\lbrace q_{i} \right\rbrace_{i=1}^{4}  $$ 
	
	\noindent where each problem is identified with an element of $\mathcal{F}_0 = f_{1} \times f_2$, the space of features of the problems in memory, all related to buying or not cellphones. Here $f_{1}$ is the space of screen sizes, $f_{2}$ is the space of RAM memory sizes:
	
	\begin{itemize} 
		\item $f_{1}= \left\lbrace 5, 5.5, 7 \right\rbrace$,
		\item $ f_{2}= \left\lbrace 16, 32\right\rbrace $,
	\end{itemize}
	
	The corresponding graph has diameter $D_{M_{\mathcal{F}_0}} = 3$.\\
	
	A case $c_i \in M$ associated to a problem $q_i$, $i=1, \ldots, 4$, is then described as:
	
	$$c_{i} = \langle q_i, a_i, r_i \rangle \in \mathcal{F}_0 \times \{0,1\} \times R$$

	\noindent where $a_i$ is the decision made (either ``buy'' or ``not buy'') while $r_i \in R$ is the result, understood as a degree of satisfaction (a real number in the interval $[0, 10]$. The cases are:\\
	\begin{align*}
		c_{1} &= \langle (5, 16), 1, 5\rangle \\
		c_{2} &= \langle (5.5, 16), 0, 10 \rangle\\
		c_{3} &= \langle (5, 32), 0, 7 \rangle \\
		c_{4} &= \langle (5.5, 32), 1, 7 \rangle \\
	\end{align*}
	
	Now suppose a new problem $p \in \mathcal{F}_0$ appears, namely to buy or not a configuration $(7, 16)$. The corresponding distances to the problems in $H(M_{\mathcal{F}_0})$ are: $d(p,q_1) = 2$, $d(p,q_2) = 1$, $d(p, q_3)= 3$ and $d(p, q_4) =2$. Then (see Figure~\ref{fig:F1}),\\
	
	\begin{align*}
		s(p, q_1) = 1 - \frac{2}{3} = \frac{1}{3}\\
		s(p, q_2) = 1 - \frac{1}{3} = \frac{2}{3} \\
		s(p, q_3) = 1 - \frac{3}{3} = 0 \\
		s(p, q_4) = 1 - \frac{2}{3} = \frac{1}{3} \\
	\end{align*}
	
	Then, the agent has to choose between $a =1$ (``buy'') and $a=0$ (``not buy''). The corresponding preferences are represented as (we assume $u(r) = r$ for every $r \in R$):\\
	
	\begin{align*}
		U(1) &= s(p,q_1) u(5) + s(p, q_4) u(7) = \\
		&=\frac{1}{3} \times 5 + \frac{1}{3} \times 7 = 4 \\
		U(0) &= s(p,q_2) u(10)  + s(p, q_3) u(7) = \\ 
		&=\frac{2}{3} \times 10 + 0 \times 7 = 6 \frac{2}{3}\\
	\end{align*}
	
	Since $U(0) > U(1)$ the decision is not buy the object.\\
\end{example}

\section{Acquiring New Information}\label{sec5}

Up to this point, we have taken the entire $M$ as source for the similarity relation. But new information may appear that could require to revise the decisions made previously and change the way of addressing new problems. There are two instances that we want to consider:

\begin{itemize}
	\item[({\bf a})] A problem $p$ that may appear with a new set of salient features $\{f_{1^p}, \ldots f_{m^p}\}$. Each $f_j$, $j=1^p, \ldots, m^p$ is either included in a space orthogonal to $\mathcal{F}$ and to each $f_k$ for $k \neq j$ or there exists a feature of $\mathcal{F}$, say $f_l$, such that $f_l \subset f_j$, i.e. $f_j$ adds extra values to feature $f_l$. 
	\item[({\bf b})] $p$ may not be compared to any other problem in $H(M)$ but only to some aspect of them.  
\end{itemize}

Case {\bf (a)} presents two subcases:

\begin{itemize}
	\item[{\bf (a1)}] A feature incorporates a new value.
	\item[{\bf (a2)}] A new feature becomes relevant. 
	
\end{itemize}

In these two instances the relevant space becomes $\mathcal{F}^p \times \prod_{k=1}^m f_{k^p}$, where $\mathcal{F}^p$ are either the new features or the ones already in $\mathcal{F}$ with the new values detected in $p$. Then, a new $s(\cdot, \cdot)$, based on the corresponding distance in the new graph, must be computed.\\

\begin{example}~\label{ex2} (Case {\bf (a1)}): consider the same $M$ as in Example~\ref{ex1}, where $H(M) = f_{1} \times f_{2}$ = $\{5, 5.5\} \times \{16, 32\}$. Recall that $M=\left\lbrace c_{1};c_{2};c_{3};c_{4}\right\rbrace$ is such that:\\
	\begin{align*}
		q_{1} &= (5, 16)\\
		q_{2} &= (5.5, 16) \\
		q_{3} &=  (5, 32) \\
		q_{4} &=  (5.5, 32) \\
	\end{align*}
	\noindent with corresponding distances: $d(q_{1},q_{2}) = 1$, $d(q_{2},q_{3}) = 2$, $d(q_{3},q_{4}) = 1$, $d(q_{1},q_{4}) = 2$, $d(q_{1},q_{3}) = 1$ and $d(q_{2},q_{4}) = 1$. Then, since the diameter of the graph is $D_{M} = 2$, the similarity function is such that: $s(q_{1},q_{2}) = \frac{1}{2}$, $s(q_{2},q_{3}) = 0$, $s(q_{3},q_{4}) = \frac{1}{2}$, $s(q_{1},q_{3}) = \frac{1}{2}$, $s(q_{2},q_{4}) = \frac{1}{2}$ and $s(q_{1},q_{4}) = 0$ (see Figure~\ref{fig:F2}).\\
	
	\begin{figure}[h!]\centering
		\includegraphics[scale=1.05]{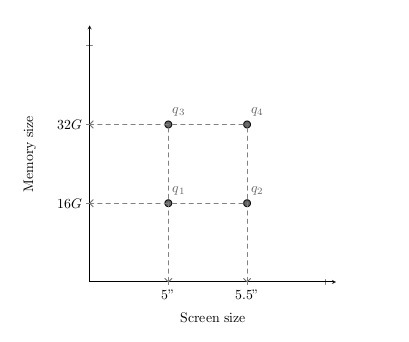}
		\caption{Example of $H(M)$ in Case {\bf (a1).}} \label{fig:F2}
	\end{figure}
	
	Now assume a new problem $p$ in which feature $f_{1}$ presents a new value, $7$. In the new graph, corresponding to $M^{\prime} = M \cup \{p\}$, while the distances between $q_1, \ldots, q_4$ are the same, the similarity function changes since the diameter of the graph is now $D_{M} = 3$. We have that: $s(q_{1},q_{2}) = \frac{2}{3}$, $s(q_{2},q_{3}) = \frac{1}{3}$, $s(q_{3},q_{4}) = \frac{2}{3}$, $s(q_{1},q_{3}) = \frac{2}{3}$, $s(q_{2},q_{4}) = \frac{2}{3}$ and $s(q_{1},q_{4}) = \frac{1}{3}$ (see, again, Figure~\ref{fig:F1}).
	
\end{example}

\begin{example}~\label{ex3} (Case {\bf (a2)}): Assume again $M$ as in Example~\ref{ex1}, but now a new feature $f_3$ becomes relevant. Then, problems $q_1, \ldots, q_4$ have to be redefined, acquiring a new component. That is, $q_i = \langle q_i^{f_1}, q_i^{f_2} \rangle$ becomes $q_i^{\prime} = \langle q_i^{f_1}, q_i^{f_2}, \bar{q}_i^{f_3} \rangle$, where $\bar{q}_i^{f_3} \in f_3$. in case $\bar{q}_i^{f_3}$ is not known with precision or it is not defined for $q_i$ (for instance, the operating system of a an old cellphone), it is assigned an arbitrary value in $f_3$. We represent this in Figure~\ref{fig:F3}, in which the redefined problems get a non-null value in the third coordinate.
	
	\begin{figure*}[t!]\centering
		\includegraphics[scale=1]{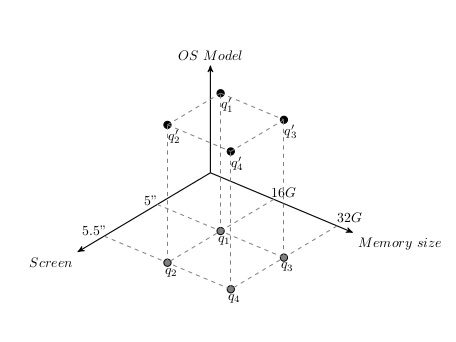}
		\caption{Example of Case {\bf (a2)}} \label{fig:F3}
	\end{figure*}

\end{example}

Case {\bf (b)} poses a different question, namely to find aspects that are shared with previous problems. It can be addressed assuming that, given a new problem $p$, it comes associated to a subspace $\mathcal{F}_0$ of $\mathcal{F}$ and a distance $\delta >0$. For each problem $q \in H(M)$ we can define $q_{|\mathcal{F}_0}$, its projection over $\mathcal{F}_0$. Then $H(M)^p = \{q \in H(M): s(p_{|\mathcal{F}_0},q_{|\mathcal{F}_0}) > \delta\}$, will be the class of problems to be taken into account for the choice of the optimal action. In case that $H(M)^p = \emptyset$, then by default we consider the entire $H(M)$.\\

\begin{example} (Case {\bf (b)}): Consider again $M$ as in Example~\ref{ex1} and the problem of buying or not a phone $p= (7, 32, 9)$, but only in terms of the comparison with the features $\mathcal{F}_0 = f_1 \times f_2$. Then, for each $q_i \in H(M)$, $q_{i_{|\mathcal{F}_0}} = q_i$ while $p_{|\mathcal{F}_0} = (7, 32)$. Figure~\ref{fig:F4}  represents this case, where the similarity function is:
	
	\begin{align*}
		s(p, q_1) = 1 - \frac{3}{3} = 0 \\
		s(p, q_2) = 1 - \frac{2}{3} = \frac{1}{3} \\
		s(p, q_3) = 1 - \frac{2}{3} = \frac{1}{3} \\
		s(p, q_4) = 1 - \frac{1}{3} = \frac{2}{3} \\
	\end{align*}
	
	If $\delta = \frac{1}{2}$, $H(M)^p = \{ q_4 \}$ and thus, since $c_4 = \langle q_4, 1, 7 \rangle$, the decision should be ``buy''.

	\begin{figure*}[h]\centering
		\includegraphics[scale=0.95]{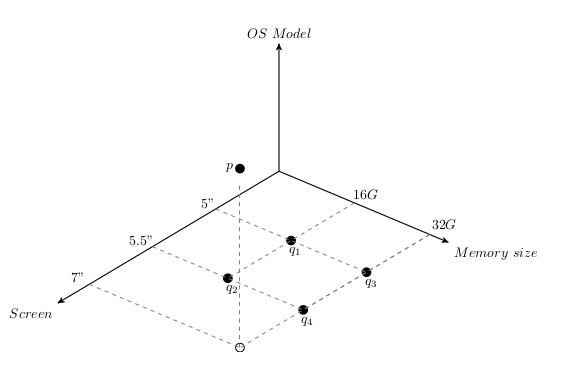}
		\caption{Example of Case {\bf (b)}} \label{fig:F4}
	\end{figure*}

\end{example}

\section{Learning}\label{sec6}

The question becomes now whether agents can anticipate new features or values of them, and consequently make choices based on those forecasts. This can only be probabilistically defined, and requires some assumptions. The first one is that all the distributions are Poisson, as to capture the intuition that the relevant events (new features and values) arise at a rate independent of the events at the previous period. This assumption can be criticized on different grounds, but without it we are forced to assume either a stronger regularity (a certain ``law'' of evolution of features and values) or a weaker one, making harder to forecast future events.

The second, and perhaps more critical, assumption is that the intervals defining the Poisson distributions represent the time interval between the formulation and solution of two different problems. While it does not need to be of a fixed duration, it tends to have a certain regularity in itself.

Let us define what it means for values and features, in terms of cases {\bf (a1)} and {\bf (a2)} (case {\bf (b)} is of different nature and will not be treated here):

\begin{itemize}
	\item[{\bf (a1)}] Any new problem $p$ has a probability $Prob(f_{j^p}) \sim Poisson(\lambda_j)$ of adding a new value to any feature $f_j$, for a given parameter $\lambda_j$. Furthermore, we assume that $\lambda_j = \lambda_k$ for any pair of features $f_j$ and $f_k$ already present in $H(M)$.
	\item[{\bf (a2)}] A new feature appears in any problem $p$ with probability $Prob(J) \sim Poisson(\lambda_J)$. 
	
\end{itemize}

The parameters $\lambda$ represent the average number of changes in either the amount of values in a feature or in the number of features per relevant interval. To determine them, we have to compute some values. Given any new problem $p$ we denote with $\Delta_{j^p}$ the amount of new values of feature $f_j$ and by $\Delta_{J^p}$ the amount of new features in $p$.

Then, given two memories $M$ and $M^{\prime}$, we define 

$$\lambda_j = \frac{\sum_{p \in M^{\prime}\setminus M} \Delta_{j^p}}{|M^{\prime} \setminus M|} \ \  \mbox{and} \ \ \lambda_J = \frac{\sum_{p \in M^{\prime}\setminus M} \Delta_{J^p}}{|M^{\prime} \setminus M|}$$

In either case the probability of $k$ new values or features is given by

$$ f(k,\lambda)=\dfrac{e^{-\lambda}\lambda^{k}}{k!} $$

The assessment of the values of $\lambda_j$ and $\lambda_J$ proceeds by progressive 
refinement, reevaluating those parameters starting from $M_0 = \emptyset$ and computing them for certain sizes of the memory. That is, every some fixed number of problems, the parameters are
calculated again.

This process of continuous updating does not necessarily converge to a fixed distributions. Agents will use the distributions they have at hand at the moment of making their decision. Furthermore, since they are anticipating events that will happen at some time in the future, they have to \emph{discount} the value of those outcomes to make meaningful comparisons with decisions with immediate results.

We can consider lotteries over a space $\mathcal{M} = \{\langle M, t,  p\rangle : M$ is a memory, $t $ is the moment at which it is assumed to be available and $p$ is a problem for which a solution is sought$\}$. Assuming preferences $\succeq_{vNM}$ over $\mathcal{M}$, satisfying the usual {\em von Neumann-Morgenstern} axioms, we can represent them by an expected utility $\bar{U}$. It is immediate that $\bar{U}(\langle M, t, p\rangle)$ corresponds to a lottery in which the probability of $\langle M, t, p\rangle$ is $1$ and the probability of every other $\langle M^{\prime}, t^{\prime}, p^{\prime}\rangle \in \mathcal{M}$ is $0$. The value of $\bar{U}(\langle M, t, p\rangle)$ can be identified with $\kappa^{t -t_0} U_{p,M}(a^*)$ where $a^*$  is the action $a$ that maximizes $\sum_{(q,a,r) \in M} s(p,q) u(r)$  and $t_0$ is the moment at which the lotteries are evaluated.\footnote{This is consistent with the two levels of inductive reasoning in learning and induction presented in \parencite{Gilboa2001, Gilboa2003}.}

\begin{example}: Consider again $M$ as in Example~\ref{ex1} and the problem of buying or not a phone $p= (7, 32, 9)$ now (period $t$) or wait until $t + 2$ to buy a better phone with a fourth feature $f_4$ and one higher value in features $f_2$ and $f_3$. Since $t+2$ is two intervals away, in terms of Poisson's distribution, the probabilities of those events are: $ f(1, 2\lambda_j)$, for $j=2,3$ and $f(1, 2 \lambda_J)$.
	
	Taking $\lambda_j = \frac{1}{2}$ and $\lambda_J = \frac{1}{20}$, we have that the probabilities of new (higher) values in $f_2$ and $f_3$ are both $\frac{1}{e}$, while that of a new feature $f_4$ is $\frac{1}{10 \ e^{10}}$.
	
	The similarities between $p$ and the problems in $H(M)$ are ($D_M = 4$):
	\begin{align*}
		s(p, q_1) = 1 - \frac{4}{4} = 0 \\
		s(p, q_2) = 1 - \frac{1}{4} = \frac{1}{4} \\
		s(p, q_3) = 1 - \frac{3}{4} = \frac{1}{4} \\
		s(p, q_4) = 1 - \frac{2}{4} = \frac{1}{2} \\
	\end{align*}
	
	While that of the potential $p^{\prime}$ are (the new $M^{\prime}$ has a diameter\footnote{In this case, the diameter is the maximal
		distance between $q_1$ and $p^{\prime}$} $D_{M^{\prime}} = 7$):
	
	\begin{align*}
		s(p^{\prime}, q_1) = 1 - \frac{7}{7} = 0 \\
		s(p^{\prime}, q_2) = 1 - \frac{6}{7} = \frac{1}{7} \\
		s(p^{\prime}, q_3) = 1 - \frac{6}{7} = \frac{1}{7} \\
		s(p^{\prime}, q_4) = 1 - \frac{5}{7} = \frac{2}{7} \\
	\end{align*}
	
	Consider the decision of buying a cellphone. We can compare two lotteries. One yields the utility of buying $p$ in $t$, denoted $U_p^t(1)$, with certainty.
	
	The other lottery yields the expected utility of buying $p^{\prime}$ in $t+2$, $U_{p^{\prime}}^{t+2}(1)$, with probability $f(1, 2\lambda_j) \times f(1, 2\lambda_j) \times f(1, 2\lambda_J) = \frac{1}{e} \times \frac{1}{e} \times \frac{1}{10 \ e^{10}} = \frac{1}{10 \ e^{12}}$.
	

	The first lottery yields:
	
	$$U_p^t(1) = s(p,q_1) \times 5 + s(p, q_4) \times 7 = \frac{7}{2} = 3 \frac{1}{2}$$
	
	\noindent while for the second lottery, we need to compute its expected utility, by taking into account the aforementioned probabilities as well as the discount rate:\footnote{Notice that it has to be applied as an \emph{interest rate}, improving the future value of the rewards.}
	
	\begin{align*}
		U_{p^{\prime}}^{t+2}(1) &= \frac{1}{10} \frac{1}{e^{12}} (s(p^{\prime}, q_1) \times \kappa \times 5 + s(p^{\prime}, q_4) \times \kappa \times 7) =\\
		&= \frac{1}{5 \ e^{12}} \ \kappa
	\end{align*}
	
	Then, the decision to postpone buying a cellphone, waiting for a potential $p^{\prime}$ brand, makes only sense if $\kappa \geq 17.5 \ e^{12}$, i.e. the future utility must be exceedingly large to compensate the extremely low chances of obtaining $p^{\prime}$ ($\sim 6 \times 10^{-7}$).
	
\end{example} 

\section{Conclusions}\label{sec7}
We have presented a graph-based definition of similarity to be used in the framework of Case-Based Decision Theory. This allows us to compute easily choices that are optimal in terms of the comparison with problems solved in the past. 

Our characterization allows to represent the acquisition of new information, defining a learning process in time. Our results indicate that while no convergence can be ensured, this allows to compare current and future decisions.

This is relevant in the context of judgment and decision-making in categorization tasks, yielding optimal answers to the intuitions and demands  for methodological advances made by \cite{newell2008cognitive}. One of the main fields in which these tasks are particularly relevant is the adoption of new technologies, specifically in the problem of purchasing gadgets with new features or for which future additions can be expected. Further work involves refining this approach and running experiments to compare with the decisions actually made by human agents.

\printbibliography

\bigskip

\appendix

\section{Proof of Theorem 1.} \label{app1}

We are going to show only that a) implies b) since the
converse is trivial. Gilboa and Schmeidler (1995) showed that Axioms 1 to 5
are equivalent to the existence of a function $s:P^{2}\rightarrow \left[ 0,1\right] $ such that for all $p\in P$, every memory $M$ with $p\notin
H=H\left( M\right) $, and every compatible $x,y\in \mathbb{R}^{H}$,
$x\succeq _{p,H}y\Longleftrightarrow \underset{q\in H}{\dsum }s\left(
p,q\right) x\left( q\right) \geq \underset{q\in H}{\dsum }s\left( p,q\right)
y\left( q\right) $.\\

Now we are going to show that considering axiom 6 the function $s\left(p,\cdot \right)$ can be transformed in a symmetric function, i.e. we are going to show that there exist a scalar $\delta _{p}$ such that we can rescale $s\left( p,\cdot \right) $, separately for each $p$, to convert it in a symmetric function. Thus, it suffices to show that there exist a function $s^{\prime }\left( p,\cdot \right) $ and a scalar $\delta _{p}$
such that:
$$s^{\prime }\left( p,q\right) =\delta _{p}s\left( p,q\right)$$

and that this function is going to verify:

$$s^{\prime }\left( p,q\right) =s^{\prime }\left( q,p\right)$$

From axiom 6,

$$x,y\in \mathbb{R}^{H^{1}},~z,w\in \mathbb{R}^{H^{2}},~l,h\in \mathbb{R}^{H^{3}}$$

\begin{itemize}
	\item \qquad \qquad $x\approx _{p,H^{1}}y\Longleftrightarrow$
\end{itemize}

\begin{multline*}
	\underset{q\in \left\{ H^{1}\backslash \left\{ m,r\right\} \right\} }{\dsum 
	}s_{H^{1}}\left( p,q\right) x\left( q\right) + s_{H^{1}}\left( p,m\right)
	x\left( m\right) + s_{H^{1}}\left( p,r\right) x\left( r\right) =\\
	\underset{q\in \left\{ H^{1}\backslash \left\{ m,r\right\} \right\} }{\dsum 
	}s_{H^{1}}\left( p,q\right) y\left( q\right) +s_{H^{1}}\left( p,m\right)
	y\left( m\right) +s_{H^{1}}\left( p,r\right) y\left( r\right)
\end{multline*}

\bigskip

\begin{itemize}
	\item \qquad \qquad $z\approx _{r,H^{2}}w \Longleftrightarrow$
\end{itemize}

\begin{multline*}
	\underset{q\in \left\{ H^{2}\backslash \left\{ m,p\right\} \right\} }{\dsum 
	}s_{H^{2}}\left( r,q\right) z\left( q\right) +s_{H^{2}}\left( r,m\right)
	z\left( m\right) + s_{H^{2}}\left( r,p\right) z\left( p\right) =\\
	\underset{q\in \left\{ H^{2}\backslash \left\{ m,p\right\} \right\} }{\dsum 
	}s_{H^{2}}\left( r,q\right) w\left( q\right) +s_{H^{2}}\left( r,m\right)
	w\left( m\right) + s_{H^{2}}\left( r,p\right) w\left( p\right)
\end{multline*}

\bigskip

\begin{itemize}
	\item \qquad \qquad $l\approx _{m,H^{3}}h \Longleftrightarrow $
\end{itemize}

\begin{multline*}
	\underset{q\in \left\{ H^{3}\backslash \left\{ p,r\right\} \right\} }{\dsum 
	}s_{H^{3}}\left( m,q\right) l\left( q\right) +s_{H^{3}}\left( m,p\right)
	l\left( p\right) +s_{H^{3}}\left( m,r\right) l\left( r\right) =\\
	\underset{q\in \left\{ H^{3}\backslash \left\{ p,r\right\} \right\} }{\dsum 
	}s_{H^{3}}\left( m,q\right) h\left( q\right) +s_{H^{3}}\left( m,p\right)
	h\left( p\right) +s_{H^{3}}\left( m,r\right) h\left( r\right)
\end{multline*}

\bigskip

\noindent and

\bigskip

\begin{itemize}
	\item \qquad \qquad $x+\alpha v_{m}^{1}\approx _{p,H^{1}}y+v_{r}^{1} \Longleftrightarrow $
\end{itemize}

\begin{multline*}
	\underset{q\in \left\{ H^{1}\backslash \left\{ m,r\right\} \right\} }{\dsum 
	}s_{H^{1}}\left( p,q\right) x\left( q\right) +s_{H^{1}}\left( p,m\right)\left[ x\left( m\right) +\alpha \right] +s_{H^{1}}\left( p,r\right) x\left(r\right) =\\
	\underset{q\in \left\{ H^{1}\backslash \left\{ m,r\right\} \right\} }{\dsum}s_{H^{1}}\left( p,q\right) y\left( q\right) +s_{H^{1}}\left( p,m\right)
	y\left( m\right) + s_{H^{1}}\left( p,r\right) \left[ y\left( r\right) +1%
	\right]
\end{multline*}

\bigskip

\begin{itemize}
	\item \qquad \qquad $z+\beta v_{m}^{2}\approx _{r,H^{2}}w+v_{p}^{2} \Longleftrightarrow $
\end{itemize}

\begin{multline*}
	\underset{q\in \left\{ H^{2}\backslash \left\{ m,p\right\} \right\} }{\dsum 
	}s_{H^{2}}\left( r,q\right) z\left( q\right) +s_{H^{2}}\left( r,m\right) %
	\left[ z\left( m\right) +\beta \right] +s_{H^{2}}\left( r,p\right) z\left(
	p\right) =\\
	\underset{q\in \left\{ H^{2}\backslash \left\{ m,p\right\} \right\} }{\dsum}s_{H^{2}}\left( r,q\right) w\left( q\right) +s_{H^{2}}\left( r,m\right)
	w\left( m\right) +s_{H^{2}}\left( r,p\right) \left[ w\left( p\right) +1\right]
\end{multline*}

\bigskip

\noindent then

$$s_{H^{1}}\left( p,m\right) \alpha =s_{H^{1}}\left( p,r\right) \ \mbox{and} \
s_{H^{2}}\left( r,m\right) \beta =s_{H^{2}}\left( r,p\right) $$

\noindent imply that

\begin{itemize}
	\item \qquad \qquad $l+\alpha v_{p}^{3}\approx _{m,H^{3}}h+\beta v_{r}^{3} \Longleftrightarrow $
\end{itemize}

\begin{multline*}
	\underset{q\in \left\{ H^{3}\backslash \left\{ p,r\right\} \right\} }{\dsum 
	}s_{H^{3}}\left( m,q\right) l\left( q\right) +s_{H^{3}}\left( m,p\right) %
	\left[ l\left( p\right) +\alpha \right] +s_{H^{3}}\left( m,r\right) l\left(
	r\right) =\\
	\underset{q\in \left\{ H^{3}\backslash \left\{ p,r\right\} \right\} }{\dsum}s_{H^{3}}\left( m,q\right) h\left( q\right) +s_{H^{3}}\left( m,p\right)
	h\left( p\right) +s_{H^{3}}\left( m,r\right) \left[ h\left( r\right) +\beta %
	\right]
\end{multline*}

\noindent then

$$s_{H^{3}}\left( m,p\right) \alpha =s_{H^{3}}\left( m,r\right) \beta $$

\noindent whenever the compared profiles are compatible.\\

Therefore, axiom 6 implies that

$$s_{H^{3}}\left( m,p\right) s_{H^{1}}\left( p,r\right) s_{H^{2}}\left(
r,m\right) =s_{H^{3}}\left( m,r\right) s_{H^{2}}\left( r,p\right)
s_{H^{1}}\left( p,m\right)$$

Using axiom 5 in the last expression we obtain:

$$s\left( m,p\right) s\left( p,r\right) s\left( r,m\right) =s\left(
m,r\right) s\left( r,p\right) s\left( p,m\right) $$

$$s\left( m,p\right) \delta _{m}=\delta _{p}s\left( p,m\right) $$

$$s^{\prime }\left( m,p\right) =s^{\prime }\left( p,m\right)$$
\begin{flushright}
	$\blacksquare$
\end{flushright}

\section{Proof of Theorem 2.} \label{app2}

We are going to show that a) implies b), the converse is trivial. From axiom 7,

\begin{itemize}
	\item \qquad \qquad $x\approx _{p,H^{1}}y \Longleftrightarrow $
\end{itemize}

\begin{multline*}
	\underset{q\in \left\{ H^{1}\backslash \left\{ m,r\right\} \right\} }{\dsum 
	}s_{H^{1}}\left( p,q\right) x\left( q\right) +s_{H^{1}}\left( p,m\right)
	x\left( m\right) +s_{H^{1}}\left( p,r\right) x\left( r\right) =\\
	\underset{q\in \left\{ H^{1}\backslash \left\{ m,r\right\} \right\} }{\dsum 
	}s_{H^{1}}\left( p,q\right) y\left( q\right) +s_{H^{1}}\left( p,m\right)
	y\left( m\right) +s_{H^{1}}\left( p,r\right) y\left( r\right)
\end{multline*}

\bigskip

\begin{itemize}
	\item \qquad \qquad $z\approx _{r,H^{2}}w \Longleftrightarrow $
\end{itemize}

\begin{multline*}
	\underset{q\in \left\{ H^{2}\backslash \left\{ m,p\right\} \right\} }{\dsum 
	}s_{H^{2}}\left( r,q\right) z\left( q\right) +s_{H^{2}}\left( r,m\right)
	z\left( m\right) +s_{H^{2}}\left( r,p\right) z\left( p\right) = \\
	\underset{q\in \left\{ H^{2}\backslash \left\{ m,p\right\} \right\} }{\dsum 
	}s_{H^{2}}\left( r,q\right) w\left( q\right) +s_{H^{2}}\left( r,m\right)
	w\left( m\right) +s_{H^{2}}\left( r,p\right) w\left( p\right)
\end{multline*}

\bigskip

\noindent and

\begin{itemize}
	\item \qquad \qquad $x+\alpha v_{m}^{1}\approx _{p,H^{1}}y+v_{r}^{1} \Longleftrightarrow $
\end{itemize}

\begin{multline*}
	\underset{q\in \left\{ H^{1}\backslash \left\{ m,r\right\} \right\} }{\dsum 
	}s_{H^{1}}\left( p,q\right) x\left( q\right) +s_{H^{1}}\left( p,m\right) %
	\left[ x\left( m\right) +\alpha \right] +s_{H^{1}}\left( p,r\right) x\left(
	r\right) = \\
	\underset{q\in \left\{ H^{1}\backslash \left\{ m,r\right\} \right\} }{\dsum 
	}s_{H^{1}}\left( p,q\right) y\left( q\right) +s_{H^{1}}\left( p,m\right)
	y\left( m\right) +s_{H^{1}}\left( p,r\right) \left[ y\left( r\right) +1%
	\right]
\end{multline*}

\bigskip

\begin{itemize}
	\item \qquad \qquad $z+\beta v_{p}^{2}\approx _{r,H^{2}}w+v_{m}^{2} \Longleftrightarrow $
\end{itemize}

\begin{multline*}
	\underset{q\in \left\{ H^{2}\backslash \left\{ m,p\right\} \right\} }{\dsum 
	}s_{H^{2}}\left( r,q\right) z\left( q\right) +s_{H^{2}}\left( r,p\right) %
	\left[ z\left( p\right) +\beta \right] +s_{H^{2}}\left( r,m\right) z\left(
	m\right) =\\
	\underset{q\in \left\{ H^{2}\backslash \left\{ m,p\right\} \right\} }{\dsum 
	}s_{H^{2}}\left( r,q\right) w\left( q\right) +s_{H^{2}}\left( r,p\right)
	w\left( p\right) +s_{H^{2}}\left( r,m\right) \left[ w\left( m\right) +1%
	\right]
\end{multline*}

\noindent then

$$s_{H^{1}}\left( p,m\right) \alpha =s_{H^{1}}\left( p,r\right) \ \mbox{and} \ 
s_{H^{2}}\left( r,p\right) \beta =s_{H^{2}}\left( r,m\right) $$

\qquad

\noindent imply that

\begin{itemize}
	\item \qquad \qquad $y+\beta v_{r}^{1}+\alpha v_{m}^{1}\preceq _{p,H^{1}}x+v_{m\text{ }}^{1} \Longleftrightarrow $
\end{itemize}

\begin{multline*}
	\underset{q\in \left\{ H^{1}\backslash \left\{ m,r\right\} \right\} }{\dsum 
	}s_{H^{1}}\left( p,q\right) y\left( q\right) +s_{H^{1}}\left( p,m\right) %
	\left[ y\left( m\right) +\alpha \right] +s_{H^{1}}\left( p,r\right) \left[
	y\left( r\right) +\beta \right] \leq \\
	\underset{q\in \left\{ H^{1}\backslash \left\{ m,r\right\} \right\} }{\dsum}s_{H^{1}}\left( p,q\right) x\left( q\right) +s_{H^{1}}\left( p,m\right) \left[ x\left( m\right) +1\right]
	+s_{H^{1}}\left( p,r\right) x\left(r\right)
\end{multline*}

\noindent then

$$s_{H^{1}}\left( p,m\right) \alpha +s_{H^{1}}\left( p,r\right) \beta \leq
s_{H^{1}}\left( p,m\right) $$

\noindent whenever the compared profiles are compatible.\\
Therefore, axiom 7 implies that

$$s_{H^{1}}\left( p,m\right) \frac{s_{H^{1}}\left( p,r\right) }{%
	s_{H^{1}}\left( p,m\right) }+s_{H^{1}}\left( p,r\right) \frac{%
	s_{H^{2}}\left( r,m\right) }{s_{H^{2}}\left( r,p\right) }\leq s_{H^{1}}\left( p,m\right)$$

Using axioms 5 and 6 we obtain:

$$s^{\prime }\left( p,m\right) \geq s^{\prime }\left( p,r\right) +s^{\prime
}\left( r,m\right) $$ 

\begin{flushright}
	$\blacksquare$
\end{flushright}

\section{Proof of Propositions.} \label{app3}

\subsection*{Proof of Proposition 1.} \label{app3.1}

The interpretation satisfies the following statements:

\begin{itemize}
	\item[$\rho ^{1})$] $p\rightarrow \bar{p}$
	
	\item[$\rho ^{2})$] $s^{\prime }\left( p,q\right) \rightarrow d\left( \bar{p},\bar{q%
	}\right) $
	
	\item[$\rho ^{3})$] $\geq _{s}\rightarrow \leq _{d}$
	
	\item[$\rho^{4})$] $u(a) \rightarrow \bar{u_{a}}$.
\end{itemize}

\noindent where $p,q \in P$, while $\bar{p}, \bar{q} \in V$ and $\bar{u_{a}} \in \mathbb{R}$.\\

$d(\cdot )$ is a metric; that is for all $\bar{p},\bar{q},\bar{m}\in V\left(
G\right) $,

\begin{itemize}
	
	\item[1.] $d\left( \bar{p},\bar{q}\right) \geq 0$, with $d\left( \bar{p},\bar{q}%
	\right) =0$\ if and only if $p=q$
	
	\item[2.] $d\left( \bar{p},\bar{q}\right) =d\left( \bar{q},\bar{p}\right) $
	
	\item[3.] $d\left( \bar{p},\bar{m}\right) \leq d\left( \bar{p},\bar{q}\right)
	+d\left( \bar{q},\bar{m}\right) $;
	
\end{itemize}

\noindent and that the similarity function satisfies

\begin{itemize}
	\item[$1^{\prime}$.] $s^{\prime }\left( p,q\right) \geq 0$
	
	\item[$2^{\prime}$.] $s^{\prime }\left( p,q\right) =s^{\prime }\left( q,p\right) $
	
	\item[$3^{\prime}$.] $s^{\prime }\left( p,m\right) \geq s^{\prime }\left( p,q\right)
	+s^{\prime }\left( q,m\right) $
\end{itemize}

Therefore, 
$$s^{\prime }\left( p,q\right) \geq s^{\prime }\left(
p,m\right) \Longrightarrow d\left( \bar{p},\bar{q}\right) \leq d\left( \bar{p%
},\bar{m}\right).$$

That is, $d$ is an interpretation of $s^{\prime}$, while $V(G)$ is an
interpretation of the problems in memory plus the current problem. Then, $%
S_{G}$ provides an interpretation of both the memory set and the similarity
relation. To see that $S_{G} \cup \Re$ yields a true interpretation of CBDT
+ Axiom 6 + Axiom 7, notice that the main aim of axioms 1 to 4 is to show
that the similarity function combined with the U-maximization is derivable
from observed preferences. Therefore, since that is settled, now we have to
show that the $d(\cdot )$ defined over G satisfies axioms 5 to 7.\\

Axiom 5 implies that, given two nonempty sets $H,H^{0}\subseteq P\backslash
\left\{ p\right\} $, and $m,q\in H,H^{0}$, then

$$\frac{s_{H}\left( p,q\right) }{s_{H}\left( p,m\right) }=\frac{%
	s_{H^{0}}\left( p,q\right) }{s_{H^{0}}\left( p,m\right) }.$$

Since $d(\cdot )$ does not depend on the history because $G$ contains all the $%
p\in P$ then trivially satisfies this axiom.\\

Axiom 6 implies that the similarity function is symmetric, i.e. that $%
s^{\prime }\left( m,p\right) =s^{\prime }\left( p,m\right) $. Since $G$ is a
connected graph then $d(\cdot )$ is a metric. Therefore, it is symmetric.\\

Axiom 7 implies that the similarity function satisfies $s^{\prime }\left(
p,m\right) \geq s^{\prime }\left( p,q\right) +s^{\prime }\left( q,m\right) $
implying that $d(\cdot )$ is a metric, since it satisfies the triangular inequality.\\

\begin{flushright}
	$\blacksquare$
\end{flushright}

\subsection*{Proof of Proposition 2.} \label{app3.2}
Notice that given a particular problem $p^{\ast }$, there exists a partition of $P$, denoted $\{C_{i}\}_{i\in N}$
(where $N$ is the set of natural numbers) in which for each $i\in N$, $%
C_{i}=\{q\in P:s^{\prime }(p^{\ast },q)=i\}$.\\

Since any theory that determines a partition of a countable set in countable classes is categorical \parencite{Keisler1977}, there exists a single model for $s^{\prime }$ up
to isomorphism. Furthermore, since there exists a single, up to linear
transformations (i.e. isomorphisms), assignation of values of cardinal
utilities into the real numbers, there exists a single model for CBDT +
Axiom 6 + Axiom 7, namely $M_{G}$.

\begin{flushright}
	$\blacksquare$
\end{flushright}


\end{document}